\documentclass[12pt]{article}
\usepackage{amssymb}
\usepackage{amsfonts}
\usepackage{amsmath}
\usepackage[mathscr]{eucal}
\usepackage{amssymb}
\usepackage{amsthm}
\usepackage{mathrsfs}
\usepackage{bbold}
\usepackage{bm}
\usepackage{graphicx}
\usepackage{caption}
\usepackage[english]{babel}
\usepackage[T2A]{fontenc}
\usepackage[utf8]{inputenc}
\usepackage{cite}
\usepackage{xcolor}

\newcommand{\be}{\begin{equation}}
\newcommand{\ee}{\end{equation}}
\newcommand{\bea}{\begin{eqnarray}}
\newcommand{\eea}{\end{eqnarray}}

\newcommand{\lb}{\label}
\newcommand{\p}[1]{(\ref{#1})}

\usepackage{MnSymbol}

\newcounter{rown}

\begin{document}
\begin{titlepage}
\vspace*{0.1cm}

\begin{center}
{\LARGE\bf On BRST Lagrangian formulation\\[.7em] of massless higher spin
fields}

\vspace{1.2cm}

{\large\bf I.L.\,Buchbinder$^{1,2,3}$\!\!,\,\,
S.A.\,Fedoruk$^1$\!,\,\,  A.P.\,Isaev$^{1,4}$\!,\,\,  V.A.\,Krykhtin$^{2}$}

\vspace{1.2cm}

\ $^1${\it Bogoliubov Laboratory of Theoretical Physics,
Joint Institute for Nuclear Research, \\
141980 Dubna, Moscow Region, Russia}, \\
{\tt buchbinder@theor.jinr.ru, fedoruk@theor.jinr.ru,
isaevap@theor.jinr.ru}

\vskip 0.5cm

\ $^2${\it Department of Theoretical Physics,
Tomsk State Pedagogical University, \\
634041 Tomsk, Russia}, \\
{\tt joseph@tspu.edu.ru, krykhtin@tspu.edu.ru}

\vskip 0.5cm

\ $^3${\it National Research Tomsk State  University,}\\{\em Lenin
Av.\ 36, 634050 Tomsk, Russia}

\vskip 0.5cm

\ $^4${\it Faculty of Physics, Lomonosov Moscow State University,
119991 Moscow, Russia}

\end{center}

\vspace{0.6cm}

\begin{abstract}
\noindent
The paper is dedicated to the blessed memory of Professor Vladislav Gavrilovich Bagrov, an
outstanding Russian scientist in the area of theoretical and mathematical physics.
He had a great influence on the formation of the scientific interests dozens
of scientists in Tomsk and Russia as a whole. Two of the authors of this paper
(I.L.B and V.A.K) are to one degree or another grateful to Professor V.G. Bagrov for
comprehensive support in the initial period of their scientific career.
Two other authors (S.A.F. and A.P.I.) are familiar with and use the work of scientists from the Tomsk School of Theoretical Physics,
founded by Professor V.G. Bagrov. The paper is devoted to certain aspects of the higher-spin
field theory, which were mainly initiated and continued during of I.L.B and V.A.K work in Tomsk.
We demonstrate in details the simplicity and clearity of
the Lagrangian formulation for free four-dimensional massless higher-spin fields within
the universal BRST approach, while describing these fields in terms of two-component
spin-tensors.
\end{abstract}

\vspace{0.8cm}

%\bigskip
\noindent PACS: 11.10.Ef, 11.30.-j, 11.30.Cp, 03.65.Pm, 02.40.Ky

\smallskip
\noindent Keywords:   higher spins, BRST construction, AdS space \\
\end{titlepage}

\setcounter{footnote}{0}
\setcounter{equation}{0}

%%%%%%%%%%%%%%%%%%%%%%%%%%%%%%%%%%%%%%%%%%%%%%
\section{Introduction}%\label{}
%%%%%%%%%%%%%%%%%%%%%%%%%%%%%%%%%%%%%%%%%%%%%%
Professor Vladislav Gavrilovich Bagrov made a significant contribution to the development
of relativistic quantum mechanics, the charged relativistic particles radiation theory,
the general theory of relativity, as well as to the development of mathematical methods of
theoretical physics. After graduating Tomsk State University his research was closely
related with Theoretical Physics Department,
Physics Faculty, Moscow Sate University where he brilliantly defended his PhD thesis.  Further his scientific and pedagogical activities
were closely connected with Tomsk State University, where  Professor V.G.\,Bagrov organized and successfully headed one of the first departments
of quantum field theory in Russia. His active scientific research and brilliant lecture courses attracted
the dozens of young scientists to Professor V.G.\,Bagrov  that led to the formation of the
Bagrov's scientific school. He also closely followed the development of the large trends
of modern physics, one of which is the higher spin field theory.

Higher spin field theory is an actively developing area of modern
theoretical and mathematical physics (see, e.g. the reviews
\cite{revVas,revSor,revBCIV,FT,RevModPhys,DidSk,reviewsV,BekSk,Snowmass,Ponomarev} and the
references therein). In the given paper we are going to discuss the
specific aspects of the BRST Lagrangian formulation for
free higher spin field theory. This approach is now well established
and published in many papers. However, many of the specific
technical aspects of the BRST approach were never presented in
details. They were given either in the form of ``it can be shown'' or
scattered among different papers. Here, for pedagogical purposes, we
want to present the key details of the calculations in one place.
Consideration is carried out on examples of free massless higher integer spin theory
in four-dimensional Minkowski and AdS spaces. The Lagrangian formulation of such theories
was given in the pioneer papers by Fronsdal \cite{Fronsdal:1978rb,Fronsdal:1978vb} and then studied
by many authors with different aims\footnote{Since our article is not a regular one, but was
prepared for a memorial collection and is a discussion of some
specific aspects of BRST approach, we deliberately intended to avoid
a giant number of literary references and refer readers to the above
reviews.}.

The BRST approach to Lagrangian construction of higher spin field theories is based on
the formulation of these theories in the extended Fock space
with vector-type creation and annihilation operators
(see e.g. \cite{BPT,Buchbinder:2006ge}). In this case, the BRST charge involves
the constraints responsible for the trace of basic fields that leads to some technical troubles especially
while derive the interacting vertices. Similar difficulties are inherent in fact all approaches
to calculating already a  cubic vertex. Therefore, in many works the constraints responsible for
the trace are not used at all. It is assumed that such  constraints can be imposed in the
final result. It is clear that this situation cannot be considered consistent from general point of view
and requires justification (see detailed discussion of this point e.g., in \cite{BKS}).

In the given paper we discuss two aspects of BRST approach. First,
we will turn attention that in four dimensions there is a way to
avoid all troubles related to trace constraints.
Extended Fock space can be built by using the Weyl-spinor creation
and annihilation operators and
higher spin fields can be realized in terms of spin-tensors with dotted and undotted
two-component indices. In this case, the trace constraints are
automatically satisfied. Secondly, we will describe the BRST
approach and Lagrangian formulation by a completely transparent
manner. Taking into account the formulation in terms of spin-tensor
fields, we will show how the Fronsdal Lagrangian is obtained within
the BRST approach after eliminating one of two  auxiliary fields.

The paper is organized as follows. In Section\,2, we briefly
describe details of the BRST Lagrangian construction for the free
massless bosonic fields in $4D$ Minkowski space in terms of
spin-tensor fields with two-component indices.
As we will see, the use of such
fields essentially simplifies the BRST construction\footnote{The
convenience of using the BRST Lagrangian formulation in terms of
spin tensor fields was demonstrated in earlier work
\cite{BuchKout}.}. Section\,3 is devoted to generalization of the
flat space BRST construction to the AdS space. In section\,4, we
explicitly derive  the Fronsdal form of Lagrangian in AdS from the
BRST Lagrangian. In Section\,5, we summarize the results obtained.

\section{Massless bosonic spin $s$ field in Minkowski space}\label{sec:mink}

As is known (see e.g. \cite{Ideas}) the bosonic integer spin $s$ field in four dimensional Minkowski space can be described in terms of the spin-tensor gauge
fields
\be
\varphi_{\alpha(s)\dot\alpha(s)}(x):=\varphi_{\alpha_1\ldots \alpha_s\dot\alpha_1\ldots\dot\alpha_s}(x)\,,
\,
\label{fi-s}
\ee
which are totally symmetric separately with respect to all undotted and all dotted indices,
$\varphi_{\alpha_1\ldots \alpha_s\dot\alpha_1\ldots\dot\alpha_s}=\varphi_{(\alpha_1\ldots \alpha_s)(\dot\alpha_1\ldots\dot\alpha_s)}$.
In order for field \eqref{fi-s} to describe a spin $s$ massless irreducible representation of the Poinc\'are group, it must satisfy the following subsidiary conditions:
\begin{eqnarray}
\partial^2\varphi_{\alpha(s)\dot\alpha(s)}(x)=0,
&\qquad&
\partial^{\dot\alpha\alpha}\varphi_{\alpha(s)\dot\alpha(s)}(x)=0.
\label{eq-1}
\end{eqnarray}

BRST Lagrangian construction is a general universal procedure for constructing a Lagrangian formulation that provides conditions (\ref{eq-1})
as consequences of the equations of motion.

The first step of the BRST Lagrangian construction is to realize equations \eqref{eq-1} in an auxiliary Fock space,
generated by the creation $c^\alpha$, $\bar{c}_{\dot\alpha}$ and annihilation $a_\alpha$, $\bar{a}^{\dot\alpha}$ operators.
They possess the following nonzero commutation relations
\be\lb{com-cr-an}
[\bar{a}^{\dot\beta},\bar{c}_{\dot\alpha}]
=\delta^{\dot\beta}_{\dot\alpha}\,,
\qquad
[a_\beta,c^\alpha]=\delta_\beta^\alpha
\ee
and act on vacuum states $|0\rangle$ and $\langle0|$, $\langle0|0\rangle=1$ as
\be\lb{vac}
\langle0|\bar{c}_{\dot\alpha}=\langle0|c^\alpha=0\,,
\qquad
\bar{a}^{\dot\alpha}|0\rangle=a_\alpha|0\rangle=0
.
\ee
Since under the Hermitian conjugation
the dotted and undotted Lorentz group representations
transform into each other,
we have for creation and annihilation operators the following standard Hermitian conjugation conditions:
\be
(a_\alpha)^\dagger=\bar{c}_{\dot\alpha}\,,\qquad
(\bar{a}_{\dot\alpha})^\dagger=c_\alpha\,.
\ee
The conjugation of vacuum states has the standard form:
\be
(|0\rangle)^\dagger=\langle0|\,.
\ee

The Fock space of the system under consideration is formed
by ket-vectors
\begin{equation}\label{GFState}
|\varphi\rangle=
\sum_{s=0}^{\infty}|\varphi_{s}\rangle
\,,\qquad
|\varphi_{s}\rangle:=\frac{1}{s!}\,\varphi_{\alpha(s)}{}^{\dot\alpha(s)}(x)\ c^{\alpha(s)}\,\bar{c}_{\dot\alpha(s)}|0\rangle
\end{equation}
and their conjugate bra-vectors
\be\label{GFState-b}
\langle\bar{\varphi}|=
\sum_{s=0}^{\infty}\langle\bar{\varphi}_s|\,,\qquad
\langle\bar{\varphi}_s|:= \frac{1}{s!}\,\langle 0|\,\bar{a}^{\dot\alpha(s)}\,a_{\alpha(s)}\ \bar{\varphi}^{\alpha(s)}{}_{\dot\alpha(s)}(x)\,,
\ee
where we have used the concise notation:
\be
c^{\alpha(s)}:=c^{\alpha_1}\ldots c^{\alpha_s}\,,
\quad
\bar{c}_{\dot\alpha(s)}:=\bar{c}_{\dot\alpha_1}
\ldots\bar{c}_{\dot\alpha_s}\,,
\qquad
a_{\alpha(s)}:=a_{\alpha_1}\ldots a_{\alpha_s}\,,
\quad
\bar{a}^{\dot\alpha(s)}:=\bar\alpha^{\dot{\alpha}_1}\ldots\bar\alpha^{\dot{\alpha}_s}\,.
\ee

States \p{GFState} and \p{GFState-b} contain equal number of operators with non-dotted and dotted indices
and, hence, their component fields have equal number of
undotted and dotted indices.
This is reflected in the fact that the states \p{GFState} and \p{GFState-b} satisfy the conditions
\be\lb{N-N0}
(N-\bar N) \,|\varphi_{s}\rangle=0 \,,
\qquad
\langle\bar{\varphi}_{s}|\,(N-\bar N)=0
\,,
\ee
 %and as a consequence
\begin{equation}\lb{N-N0-a}
(N-\bar N) \,|\varphi\rangle=0 \,,
\qquad
\langle\bar{\varphi}|\,(N-\bar N)=0
\,,
\end{equation}
where the operators
\begin{equation}
N:=c^\alpha a_\alpha\,,
\qquad
\bar{N}:=\bar{c}_{\dot\alpha}\bar{a}^{\dot\alpha}\,,
\qquad N^\dagger=\bar N\,, \lb{ex-N}
\end{equation}
are the particle number operators. Since
 the states $|\varphi_{s}\rangle$ and $\langle\bar{\varphi}_{s}|$ contain $2s$ creation and annihilation operators,
 we have equations
\be\lb{N-N0s}
\frac12\left(N+\bar N\right) |\varphi_{s}\rangle=s\,|\varphi_{s}\rangle \,,
\qquad
\frac12\,\langle\bar{\varphi}_{s}|\left(N+\bar N\right)=s\,\langle\bar{\varphi}_{s}|
\,.
\ee

The vectors $|\varphi_{s}\rangle$ (or $\langle\bar{\varphi}_{s}|)$ describe massless integer spin $s$ field
if the component field $\varphi_{\alpha(s)}{}^{\dot\alpha(s)}(x)$ satisfies the subsidiary conditions
(\ref{eq-1}). These conditions can be reformulated in terms of operators acting in the Fock space. For this aim,
let us introduce the operators
\begin{eqnarray}
\lb{op-0-0}
\ell_0 &:=& \partial^2\,,
\\ [6pt]
\lb{op-1-0}
\ell &:=&(a\sigma^m\bar{a})\,\partial_m\,,
\\ [6pt]
\label{op-t1-0}
\ell^{+}&:=&-(c\sigma^m\bar{c})\,\partial_m\,,
\end{eqnarray}
where
\begin{eqnarray}
(a\sigma^m\bar{a})\ := a^\alpha\sigma^m_{\alpha\dot\alpha}\bar{a}^{\dot\alpha}
\,,\qquad
(c\sigma^m\bar{c})\ := c^\alpha\sigma^m_{\alpha\dot\alpha}\bar{c}^{\dot\alpha}\,
\end{eqnarray}
and $\sigma^m_{\alpha\dot\alpha}$ are the invariant matrices on Lorentz group (see e.g. \cite{Ideas}). Then, it is easy to show that the
following constraints
\be
\ell_0|\varphi\rangle=
\ell|\varphi\rangle= 0
\label{eq-1F}
\ee
 are equivalent to the conditions \eqref{eq-1} for
 each component \eqref{fi-s} of \eqref{GFState}.
As a result, the Fock space vectors $|\varphi_{s}\rangle$ under constraints (\ref{eq-1F}) can be treated as the irreducible representation
of the irreducible massless integer spin $s$ representation of the Poinc\'are group.

The second step of the BRST Lagrangian construction is finding the commutator algebra of operators \p{op-0-0}--\p{op-t1-0}.
It is easy to see that these operators form a closed algebra, where
the only non-zero commutator has the form
\begin{equation}\label{alg(0)}%\label{algebra}
[\ell^{+},\ell]=K\,\ell_0\,,
\end{equation}
where the operator $K$ is
\begin{equation}
K:=N +\bar{N}+2\,.
\lb{ex-K}
\end{equation}

The third step of the Lagrangian formulation in the approach under consideration is applying  the BRST construction
in deriving the BRST Lagrangian.
We introduce the BRST charge, assuming that
the operators \p{op-0-0}--\p{op-t1-0} are first class constraints
of still unknown Lagrangian gauge theory. Our aim is to find such a gauge theory.

According to the standard definition, the BRST charge acts in an extended Fock space, including, in addition to oscillators $c,\, \bar{c},\, a,\, \bar{a}$, the fermion ghost “coordinates” $\eta_{0},\,\eta,\,{\eta } ^{+}$ and the corresponding ghost ``momenta'' $\mathcal{P}_{0},\,\mathcal{P}^{+},\,\mathcal{P}$; under Hermitian conjugation we
have conditions $\eta^ {+}_{0}=\eta_{0}$ and $\mathcal{P}^{+}_{0}=\mathcal{P}_{0}$.
The only non-zero  anticommutation relations for these operators have the form
\be
\{\eta,\mathcal{P}^+\}
 \ = \ \{\mathcal{P}, \eta^+\}
 \ = \ \{\eta_0,\mathcal{P}_0\}
 \ = \ 1.
\label{ghosts}
\ee
The ghost operators act on the vacuum state $|0\rangle$ by the rule
\be \lb{ghost-vac}
\eta|0\rangle \ = \ \mathcal{P}|0\rangle \ = \ \mathcal{P}_0|0\rangle \ = \ 0\,.
\ee
All the ghost operators possess the standard  ghost numbers,
$gh(\eta_0)= gh(\eta)=gh(\eta^+)=1$ and $gh(\mathcal{P}_0)=gh(\mathcal{P})=gh(\mathcal{P}^+)=-1$.

The BRST charge is written as follows
\be\label{Q(0)}
Q \ = \ \eta_0\ell_0 \ + \ \eta^+ \ell \ + \ \eta\,\ell^{+} \ + \
K\eta^+\eta\, \mathcal{P}_0 \
\ee
with ghost number $gh(Q)$ = $1$.
The operator \p{Q(0)} acts in the extended Fock space of the vectors
\be
|\Phi\rangle \ = \
|\varphi\rangle \ + \ \eta_0\mathcal{P}^+|\varphi_1\rangle \ + \ \eta^+\mathcal{P}^+|\varphi_2\rangle
\label{extened vector}
\ee
which have a zero ghost number. It can be checked that the operator $Q$ (\ref{Q(0)}) is nilpotent by definition, $Q^2=0$.

To get a correct description of the system under consideration, we assume that the extended Fock space formed by states \p{extened vector} must be restricted by additional conditions similar to the conditions \p{N-N0}, \p{N-N0-a} and \p{N-N0s} in the initial Fock space.
Operators for these additional conditions must commute with the BRST charge \p{Q(0)}.
One of such additional
operators is the operator $N-\bar{N}$ in conditions \p{N-N0} and \p{N-N0-a}:
$[N-\bar{N},Q]=0$.
The second condition is defined by the operator
\be
S=\frac{1}{2}\left( N+\bar{N}\right)+\eta^+\mathcal{P}+\mathcal{P}^+\eta\,,\qquad
[S,Q]=0\,,
\label{op-NS}
\ee
which is the BRST extension of the operator in conditions  \p{N-N0s}.

Since the component states are zero eigenvectors of  the operator $N-\bar N$:
\begin{equation}
(N-\bar N) \,|\varphi\rangle=0 \,,\quad (N-\bar N) \,|\varphi_{1,2}\rangle=0 \,,
\qquad
\langle\bar{\varphi}|\,(N-\bar N)=0
\,,\quad
\langle\bar{\varphi}_{1,2}|\,(N-\bar N)=0 \,,
\end{equation}
the vector \eqref{extened vector} is also zero-eigenvector of this operator:
\be\lb{N-N}
(N-\bar N) \,|\Phi\rangle=0 \,,
\qquad
\langle\bar{\Phi}|\,(N-\bar N)=0
\ee
Then, it is convenient to use the expansion of  vector \eqref{extened vector} in terms of eigenvectors $|\Phi_s\rangle$
of operator \p{op-NS}:
%has the form
\be
\label{Phi-s}
|\Phi\rangle
=
\sum_{s=0}^\infty|\Phi_s\rangle
\,,
\qquad
|\Phi_s\rangle=
|\varphi_s\rangle \ + \ \eta_0\mathcal{P}^+|\varphi_{1(s-1)}\rangle \ + \ \eta^+\mathcal{P}^+|\varphi_{2(s-2)}\rangle
\,,
\ee
where vectors $|\Phi_s\rangle$ satisfy the condition
\be\lb{S-eq}
S\,|\Phi_s\rangle=s\,|\Phi_s\rangle \,.
\ee
Therefore, we can consider separately each spin $s$ field theory which is described by the state $|\Phi_s\rangle$.

The fourth step of the BRST Lagrangian formulation is
finding the equation of motion in the extended Fock space. The main requirement for
obtaining such an equation of motion is that this equation should, as a consequence, reproduce the conditions
for an irreducible representation in terms of the Fock space (\ref{eq-1F}).
Based on this requirement, we postulate the equation of motion in the form
\begin{eqnarray}\label{eqQ}
Q\,|\Phi_s\rangle \ = \ 0.
\end{eqnarray}
Due to the nilpotency of BRST charge $Q$, we see that the vector $|\Phi_s\rangle$ in equation (\ref{eqQ}) is defined up to gauge transformation
\begin{eqnarray}
|\Phi'_s\rangle \ = \ |\Phi_s\rangle \ + \ Q\,|\Lambda_s\rangle \,,
\label{gauge transf}
\end{eqnarray}
where $|\Lambda_s\rangle$ is the extended Fock space valued gauge parameter of the form
\begin{eqnarray}\label{gtQ}
|\Lambda_s\rangle \ = \ \mathcal{P}^+|\lambda_{s-1}\rangle\,,
\label{gauge parameter}
\end{eqnarray}
that has a ghost number equal to $-1$.
The fields $|\varphi_1\rangle$, $|\varphi_2\rangle$ and the gauge parameter  $|\lambda\rangle$
in relations (\ref{extened vector}),
(\ref{gauge parameter}) have a similar decomposition in extended Fock space like
$|\varphi\rangle$ in \eqref{GFState}.

The equation of motion \p{eqQ} can be equivalently rewritten as system of equations for the component vectors $|\varphi_s\rangle,\, |\varphi_{1(s-1)}\rangle,\, |\varphi_{2(s-2)}\rangle$, entering extended vector
\p{extened vector} for given $s$, in the form
\begin{eqnarray}
&&
\ell_0|\varphi_s\rangle-\ell^+|\varphi_{1(s-1)}\rangle\,,
=0
\label{236}
\\[5pt]
&&{}
\ell\,|\varphi_s\rangle-\ell^+|\varphi_{2(s-2)}\rangle+K|\varphi_{1(s-1)}\rangle
=0\,,
\label{237}
\\ [5pt]
&&{}
\ell_0|\varphi_{2(s-2)}\rangle-\ell\,|\varphi_{1(s-1)}\rangle
=0.
\label{238}
\end{eqnarray}
The gauge transformation \p{gauge transf} yields the following system of gauge transformations for the component vectors:
\be
\delta|\varphi_s\rangle \ = \ \ell^+\,|\lambda_{s-1}\rangle\,,
\qquad
\delta|\varphi_{1(s-1)}\rangle \ = \ \ell_0\,|\lambda_{s-1}\rangle\,,
\qquad
\delta|\varphi_{2(s-2)}\rangle \ = \ \ell\,|\lambda_{s-1}\rangle.
\label{GTAdS0}
\ee

Let us show that the equations \p{236},\, \p{237},\, \p{238} together with gauge transformations \p{GTAdS0} reproduce the conditions of irreducible representation (\ref{eq-1F}) in terms of Fock space (\ref{eq-1F}).
First, we remove away the vector $|\varphi_{1(s-1)}\rangle$ using gauge transformations \eqref{GTAdS0}. After this, we move to residual gauge transformations \eqref{GTAdS0} with gauge parameter $|\lambda_{s-1}\rangle$ restricted by $\ell_0\,|\lambda_{s-1}\rangle=0$. Note that equation of motion \eqref{238} after the partial gauge fixing $|\varphi_{1(s-1)}\rangle=0$ takes the form $\ell_0|\varphi_{2(s-2)}\rangle=0$. Since $[\ell,\ell_0]=0,$ this equation is invariant under the transformation $\delta|\varphi_{2(s-2)}\rangle \ = \ \ell\,|\lambda_{s-1}\rangle$ with restricted gauge parameter $|\lambda_{s-1}\rangle.$ This on-shell gauge transformation allows us
to remove away the vector $|\varphi_{2(s-2)}\rangle$. As a result we arrive at conditions  $\ell_0|\varphi_{s}\rangle =0,\, \ell|\varphi_{s}\rangle=0.$ Hence, we have shown that the equation of motion \p{eqQ} correctly reproduces the conditions of irreducible representations (\ref{eq-1F}).
However, after eliminating all the auxiliary  fields in this way,
we lose the Lagrangian formulation which, as well known, is impossible for higher spin theories without the auxiliary fields.

The Lagrangian corresponding to the equation of motion (\ref{eqQ}) is written
as follows
\be
{\cal L}_s
\ = \
\int d\eta_0\; \langle\bar\Phi_s|\,Q\,|\Phi_s\rangle
\,.
\label{actionQ}
\ee
It is evident that the Lagrangian is gauge invariant under the transformation (\ref{gauge transf}) by construction.

Substituting BRST operator \eqref{Q(0)} into \eqref{actionQ} we rewrite identically the Lagrangian
(\ref{actionQ}) in terms of component vectors the form
\begin{eqnarray}
\mathcal{L}_s
&=&
\langle\bar\varphi_s|\Bigl\{
\ell_0|\varphi_s\rangle-\ell^+|\varphi_{1(s-1)}\rangle
\Bigr\}
%\nonumber
%\\
%&&{}
-\langle\bar\varphi_{1(s-1)}|\Bigl\{
\ell\,|\varphi_s\rangle-\ell^+|\varphi_{2(s-2)}\rangle+K|\varphi_{1(s-1)}\rangle
\Bigr\}
\nonumber
\\ [5pt]
&&{}
-\langle\bar\varphi_{2(s-2)}|\Bigl\{
\ell_0|\varphi_{2(s-2)}\rangle-\ell\,|\varphi_{1(s-1)}\rangle
\Bigr\}.
\label{LagrAdS0}
\end{eqnarray}
As a result we see that the general description of the free massless higher integer spin $s$ field theory is given in terms of triplet of spin-tensor field
$|\varphi_s\rangle$ and the auxiliary spin-tensor fields $|\varphi_{1(s-1)}\rangle ,\, |\varphi_{2(s-2)}\rangle.$ All the fields are traceless by definition\footnote{Triplet structure of the higher integer spin field Lagrangian can also be derived from tensionless limit of string theory \cite{ST}.}.

Different possible ways of partially eliminating the auxiliary
fields and/or partially fixing the gauge can lead to formally
different but equivalent forms of writing the Lagrangian under
consideration. For example, the  Lagrangian \eqref{LagrAdS0} allows
us to derive simply enough the Fronsdal Lagrangian
\cite{Fronsdal:1978rb}. It is obtained if we remove the vector
$|\varphi_{1(s-1)}\rangle$ from Lagrangian \eqref{LagrAdS0} with
help of its equation of motion
\eqref{237}. Then the Lagrangian is rewritten in
terms of two traceless fields $|\varphi_{s}\rangle$ and
$|\varphi_{2(s-2)}\rangle$. These two fields can be combined in one double traceless field (see \eqref{Phi} below)
that finally leads to the Fronsdal Lagrangian
\cite{Fronsdal:1978rb}. In the next section we will discuss a
generalization of BRST approach to construct the Lagrangian
formulation for free massless spin $s$ field in the $AdS_4$ space
and describe in details the derivation of the corresponding Fronsdal
Lagrangian. However, we can express the field
$|\varphi_{2(s-2)}\rangle$ from the equation of motion \p{238} in
the form $|\varphi_{2(s-2)}\rangle =
\frac{1}{\partial^2}\ell\,|\varphi_{1(s-1)}\rangle$, where we have
used that $\ell_0 = \partial^2$ and substitute into \p{LagrAdS0}. It
yields the non-local Lagrangian in terms of fields
$|\varphi_{s}\rangle$ and $|\varphi_{1(s-1)}\rangle.$ Such a form of
the Lagrangian for free massless higher integer spin field was
considered e.g. in \cite{FS}.

\setcounter{equation}{0}

\section{Massless bosonic spin $s$ field in $AdS_4$}\label{sec:ads0}
Let us
consider a generalization of the flat space Lagrangian formulation
developed in the previous section to curved space.

The derivation of the Lagrangian formulation in the AdS space is
carried out in the same way as was done in the previous section.
Therefore, here we focus only on essentially new aspects. Since the
main element of the formulation is the algebra of first-class
constraints, we first need to generalize the constraint operators
\p{op-0-0}--\eqref{op-t1-0} and derive their algebra. The curved
space is described by the vielbein $e_{\mu}{}^m(x)$ and
spin-connection $\omega_\mu{}^{mn}(x)=-\omega_\mu{}^{nm}(x)$, where
the Greek indices $\mu$, $\nu$ are the curved space ones and the
Latin indices $m$, $n$  are the tangent space ones.  The indices
$m$, $n$ are lowered and raised by flat metrics $\eta_{mn}$ and
$\eta^{mn}$.

We consider the case of a covariantly constant vielbein that allows us to express the Christoffel symbols in a standard way in terms of the vielbein and spin connection. As in the previous section, we will construct Lagrangian formulation in curved space using the Fock formalism. Therefore we need to define the basic differential geometric of objects in terms of Fock space vectors \p{GFState}. The covariant derivative operator $D_\mu$ acting on Fock space vectors is written as follows
\be\lb{D-t}
D_\mu\ =\ \partial_\mu \ + \ \frac{1}{2}\ \omega_\mu{}^{mn}\,\mathcal{M}_{mn}\,,
\ee
where\footnote{We use the notation
$\sigma_{mn}=-\frac{1}{4}(\sigma_{m}\bar{\sigma}_{n}-\sigma_{n}\bar{\sigma}_{m})$,
$\tilde{\sigma}_{mn}=-\frac{1}{4}(\tilde{\sigma}_{m}\sigma_{n}-\tilde{\sigma}_{n}\sigma_{m})$.}
\be\lb{M-op}
\mathcal{M}_{mn} \ = \  c^\alpha \,(\sigma_{mn})_\alpha{}^\beta\,a_\beta \ + \
\bar{c}_{\dot{\alpha}}\,(\tilde{\sigma}_{mn})^{\dot{\alpha}}{}_{\dot{\beta}}\,\bar{a}^{\dot{\beta}}
\ee
are the Lorentz algebra generators obeying the commutation relations
\be
[\mathcal{M}_{mn},\mathcal{M}_{kl}]=\eta_{ml}\mathcal{M}_{nk}+\eta_{nk}\mathcal{M}_{ml}-\eta_{mk}\mathcal{M}_{nl}-\eta_{nl}\mathcal{M}_{mk}\,.
\ee
Commutators among the operators \p{D-t} have the form
\begin{equation}
\label{D-mu-alg}
\left[ D_\mu ,D_\nu \right] \ = \ \frac{1}{2}\,R_{\mu\nu}{}^{mn}\mathcal{M}_{mn}\,,
\end{equation}
where the curvature tensor is
\be
R_{\mu\nu}{}^{m}{}_{n} \ = \ \partial_\mu\omega_\nu{}^{m}{}_n -\partial_\nu\omega_\mu{}^{m}{}_n
+[\omega_\mu,\omega_\nu]^{m}{}_n\,.
\ee
In the AdS space, the curvature tensor has the form
\be
\lb{cur-2}
R_{\mu\nu}{}^{\lambda\rho} \ = \ R_{\mu\nu}{}^{mn}e^{\lambda}{}_m e^{\rho}{}_n
\ = \
\kappa\,(\delta_\mu^\lambda\delta_\nu^\rho-\delta_\mu^\rho\delta_\nu^\lambda)
\,,
\ee
with a negative constant,  $\kappa<0$.

As a natural generalization  of operators \p{op-1-0},
\p{op-t1-0} we take
\begin{eqnarray}
\lb{op-1}
l &:=& (a\sigma^m\bar{a})\,e^\mu{}_m D_\mu\,,
\\ [6pt] \lb{op-t1}
l^+&:=& -(c\sigma^m\bar{c})\,e^\mu{}_m D_\mu\,.
\end{eqnarray}
Commutator of operators (\ref{op-1}) and \p{op-t1}
\be\lb{com-ll-6}
\left[l^+,l \right] \ = \ K\;l_0
\ee
yields the
operator
\be\lb{op-0}
l_0 := D^2 \ + \
\kappa\left(N\bar{N}+N+\bar{N}\right)\,,
\ee
where \be\lb{D2}
D^2 \
:=\ g^{\mu\nu}\left(D_\mu D_\nu-\Gamma_{\mu\nu}^\lambda
D_\lambda\right) \ =\ \frac{1}{\sqrt{-g}}\, D_\mu
\sqrt{-g}g^{\mu\nu}D_\nu
\ee
and, as usual, $g=\det g_{\mu\nu}$,
$g_{\mu\nu}=e_{\mu}{}^m e_{\nu}{}_m$.
Operator $K$ is defined in eq. \p{ex-K}.
The operator $D^2$ is
a higher-spin generalization of the Laplace-Beltrami operator whereas the
operator \p{op-0} is a generalization of the flat space operator
\p{op-0-0} to the AdS space.

One can show by direct calculations that the operators \p{op-1},
\p{op-t1}, \p{op-0} form a closed algebra: in addition to \p{com-ll-6}
the remaining commutators of this algebra are
\be\label{algebra-r}
[l,l_0]=2\kappa\,(K+1)\,l\,,\qquad [l_0,l^+]=2\kappa\,(K-1)\,l^+\,.
\ee
The BRST operator constructed on the base of this algebra has
the form
\begin{eqnarray}
Q&=&
\eta_0l_0+\eta^+l+\eta\, l^++K\eta^+\eta\mathcal{P}_0
\nonumber
\\[0.5em]
&&{}
-2\kappa(K-1)\eta_0\eta \mathcal{P}^+
+2\kappa(K+1)\eta_0\eta^+\mathcal{P}
-4\kappa\, \eta_0\eta^+\eta\,\mathcal{P}^+\mathcal{P}
\,,
\label{QAdS}
\end{eqnarray}
where we have introduced
the fermionic ghost operators $\eta_0$, $\eta$, $\eta^+$ and $\mathcal{P}_0$, $\mathcal{P}^+$, $\mathcal{P}$,
which have the anticommutation relations \p{ghosts} and act on the vacuum as in \p{ghost-vac}.
The operator $Q$ \p{QAdS} is nilpotent by construction and acts  in the extended Fock space of the vectors\footnote{We keep the same notation for the vector $|\Phi_s\rangle$ in the AdS space. This should not
lead to misunderstandings.}
\be
\label{Phi-s-AdS}
|\Phi_s\rangle=
|\phi_s\rangle \ + \ \eta_0\mathcal{P}^+|\phi_{1(s-1)}\rangle \ + \ \eta^+\mathcal{P}^+|\phi_{2(s-2)}\rangle
\,.
\ee
As
in the case of flat space, in AdS space, the operators  $N-\bar
N$ and $S$ (see \eqref{op-NS}) commute with BRST operator
\eqref{QAdS}.  Therefore, the vector \p{Phi-s-AdS} is the
eigenvector of these operators with the eigenvalues $0$ and $s$,
respectively (see \p{N-N} and \p{S-eq}). The components
$|\phi_s\rangle$, $|\phi_{1(s-1)}\rangle$, $|\phi_{2(s-2)}\rangle$
have the form \p{GFState}, where the numbers of dotted and undotted
indices equal $s$, $s-1$ and $s-2$, respectively.

The further procedure for finding the BRST description in the AdS
space is similar to consideration in the previous section for the
flat case. In particular, the BRST equations of motion is \p{eqQ},
where now the BRST operator is \eqref{QAdS}. This equation is gauge
invariant (see \p{gauge transf}, \p{gauge parameter}) that allows
us to
show that the equation of motion reproduces the conditions
\be
\bigl[l_0-2\kappa(K-1)\bigr]|\phi_s\rangle =0,\,
\quad
l|\phi_s\rangle =0\,,
\label{AdS-ir}
\ee
which are used in definition of
the irreducible massless spin $s$ representation of the AdS group.

The off-shell Lagrangian
that
yields the equation of motion
$Q|\Phi_s\rangle =0$ still has the form \eqref{actionQ}, where the BRST
charge is given by \p{QAdS}. After integration with respect the ghost
variables we obtain the  Lagrangian in terms of component vectors
$|\phi_{s}\rangle$, $|\phi_{2(s-1)}\rangle$ and
$|\phi_{1(s-2)}\rangle$ in the form
\begin{eqnarray}
\mathcal{L}_s
&=&
\langle\bar\phi_s|\Bigl\{
\bigl[l_0-2\kappa(K-1)\bigr]|\phi_s\rangle-l^+|\phi_{1(s-1)}\rangle
\Bigr\}
\nonumber
\\
&&{}
-\langle\bar\phi_{1(s-1)}|\Bigl\{
l\,|\phi_s\rangle-l^+|\phi_{2(s-2)}\rangle+K|\phi_{1(s-1)}\rangle
\Bigr\}
\nonumber
\\
&&{}
-\langle\bar\phi_{2(s-2)}|\Bigl\{
\bigl[l_0+2\kappa(K+1)\bigr]|\phi_{2(s-2)}\rangle-l\,|\phi_{1(s-1)}\rangle
\Bigr\}.
\label{LagrAdS}
\end{eqnarray}
By construction, the Lagrangian is invariant under the
gauge transformations
\be
\delta|\phi_s\rangle=l_1^+|\lambda_{s-1}\rangle\,,
\qquad
\delta|\phi_{1(s-1)}\rangle=l_0|\lambda_{s-1}\rangle\,,
\qquad
\delta|\phi_{2(s-2)}\rangle=l_1|\lambda_{s-1}\rangle\,.
\label{GTAdS}
\ee

The Lagrangian is described by the triplet of the fields
$|\phi_{s}\rangle$, $|\phi_{2(s-1)}\rangle$ and
$|\phi_{1(s-2)}\rangle$. As in the flat space, we can remove the
auxiliary vector $|\phi_{1(s-1)}\rangle$ by its equation of motion.
Then we obtain the Lagrangian in terms two vectors
$|\phi_{s}\rangle$ and $|\phi_{2(s-2)}\rangle$, which corresponds to
two traceless fields that can be combined into one double traceless
field. In this case we automatically get the Lagrangian in the
Fronsdal form. Also it is evident that in flat limit, the Lagrangian
\p{LagrAdS} coincides with the Lagrangian (\ref{actionQ}).

\section{Fronsdal form of Lagrangian in $AdS_4$}\label{sec:fronsd}

As we pointed out, the Lagrangian of a free massless integer spin field can be written in different forms. Its most commonly used form in the literature
is the Fronsdal Lagrangian \cite{Fronsdal:1978vb}.
In this section we will demonstrate in detail how the Fronsdal Lagrangian is
obtained from general Lagrangian \p{LagrAdS}.

First of all we convert the spinor indices in the component fields of the vectors
\p{GFState} and \p{GFState-b} into the vector indices
\begin{equation}
\phi^{\,\mu(s)} :=
\phi^{\,\mu_1\ldots \mu_s}
=
\frac{(-1)^s}{2^s}\;
{\sigma}^{\,\mu_1}_{\alpha_1\dot\beta_1}\ldots {\sigma}^{\,\mu_s}_{\alpha_s\dot\beta_s}
\;\phi^{\,\alpha(s)\dot\beta(s)} \,.
\label{spin2tens}
\end{equation}
These tensor fields $\phi_{\mu_1\ldots \mu_s}$ obtained from the spin-tensor fields $\phi_{\alpha(s)\dot\beta(s)}$
by \eqref{spin2tens}, are
automatically
symmetric, $\phi_{\,\mu_1\ldots \mu_s}=\phi_{\,(\mu_1\ldots \mu_s)}$,
and  traceless, $g^{\mu_1\mu_2}\phi_{\mu_1\mu_2\ldots \mu_s}=0$.

Now, eliminating the dependence on the annihilation and creation operators by using \p{com-cr-an}, \p{vac}
and passing in the Lagrangian \eqref{LagrAdS} from the Fock space vectors to the usual tensor fields, we
obtain the Lagrangian in terms of conventional tensor fields:
\begin{eqnarray}
\mathcal{L}_s
&=&
\phi^{\mu(s)}\Bigl\{
\bigl[\nabla^2+\kappa(s^2-2s-2)\bigr]\phi_{\mu(s)}-s\nabla_\mu\phi_{1\mu(s-1)}
\Bigr\}
\nonumber
\\
&&{}
+\phi_1^{\mu(s-1)}\Bigl\{
s\nabla^\nu\phi_{\nu\mu(s-1)}
+\frac{s-1}{2}\;\nabla_\mu \phi_{2\,\mu(s-2)}
-s\phi_{1\,\mu(s-1)}
\Bigr\}
\nonumber
\\
&&{}
-\frac{1}{4}\;\phi_2^{\mu(s-2)}\Bigl\{
\bigl[\nabla^2+\kappa(s^2+2s-2)\bigr]\phi_{2\mu(s-2)}
+2(s-1)\nabla^\nu \phi_{1\,\nu\mu(s-2)}
\Bigr\}
\,.
\label{LagrAdS-t}
\end{eqnarray}
Gauge transformations \eqref{GTAdS} can be rewritten in terms of the above tensor fields in the Lagrangian
\p{LagrAdS-t} and have the form:
\be \label{GTAdS-t1}
\begin{array}{rcl}
\delta\phi_{\mu(s)}&=&\displaystyle s\,\nabla_{\mu}\lambda_{\mu(s-1)}
-\frac{(s-1)}{2}\,g_{\mu(2)}\,\nabla^\nu\lambda_{\nu\mu(s-2)}
\,,
\\[7pt]
\delta\phi_{1\,\mu(s-1)}&=&\displaystyle \bigl[\nabla^2+\kappa(s^2-1)\bigr]\,\lambda_{\mu(s-1)}
\,,
\\[7pt]
\delta\phi_{2\,\mu(s-2)}&=&{}\displaystyle -2(s-1)\,\nabla^{\nu}\lambda_{\nu\mu(s-2)}
\,.
\end{array}
\ee
Note that Lagrangian \p{LagrAdS-t} and gauge transformations \p{GTAdS-t1} are completely equivalent to
initial Lagrangian \p{LagrAdS} and corresponding gauge transformations written in terms of the Fock space.
Lagrangian \p{LagrAdS-t} depends on triplet of fields: basic field $\phi_{\mu(s)}$ and two
auxiliary fields $\phi_{1\mu(s-1)}$ and $\phi_{2\mu(s-2)}$.

Passage to the Fronsdal Lagrangian is carried out as follows. The auxiliary field $\phi_{1\,\mu(s-1)}$ in
Lagrangian \p{LagrAdS-t} obeys the algebraic equation of motion
\be\lb{aux-f-expr}
\phi_{1\,\mu(s-1)}
=
\nabla^\nu\phi_{\nu\mu(s-1)}
+\frac{s-1}{2s}\;\nabla_\mu \phi_{2\,\mu(s-2)}
-
\frac{s-2}{4s}\;g_{\mu(2)}\nabla^\rho\phi_{2\rho\mu(s-3)}
\ee
that allows
us
to express this field in terms of the other fields  $\phi_{\mu(s)}$ and $\phi_{2\mu(s-2)}$.
Substituting the expression \p{aux-f-expr} into \p{LagrAdS-t}, we obtain the Lagrangian
\begin{eqnarray}
\mathcal{L}_s
&=&
\phi^{\mu(s)}\Bigl\{
\bigl[\nabla^2+\kappa(s^2-2s-2)\bigr]\phi_{\mu(s)}
\nonumber
\\
&&\hspace{10ex}{}
-s\nabla_\mu\nabla^\nu\phi_{\nu\mu(s-1)}
-\frac{s-1}{2}\;\nabla_\mu\nabla_\mu \phi_{2\mu(s-2)}
\Bigr\}
\nonumber
\\[.5em]
&&{}
-\frac{1}{4}\;\phi_2^{\mu(s-2)}\Bigl\{
\Bigl[\frac{2s-1}{s}\nabla^2+\kappa(s^2+2s-2)\Bigr]\phi_{2\mu(s-2)}
+2(s-1)\nabla^\nu\nabla^\rho \phi_{\nu\rho\mu(s-2)}
\nonumber
\\
&&\hspace{10ex}{}
+\frac{(s-1)(s-2)}{s}\nabla^\nu\nabla_\mu \phi_{2\nu\mu(s-3)}
-\frac{s-2}{s}\nabla_\mu\nabla^\nu \phi_{2\nu\mu(s-3)}
\Bigr\}
\label{LagrF2}
\end{eqnarray}
in terms of two traceless fields $\phi_{\mu(s)}$ and $\phi_{2\mu(s-2)}$.

Now we combine these two traceless fields into one double traceless field $ {h}_{\mu(s)}$ as follows:
\be\label{Phi}
{h}_{\mu(s)}
\ = \
\phi_{\mu(s)} -\frac{1}{4}\,g_{\mu(2)}\phi_{2\mu(s-2)}\,.
\ee
In this definition,
the field $\phi_{2\mu(s-2)}$ fixes\footnote{Here we use notations for the trace part as in \cite{Fronsdal:1978vb}.}
the trace part ${h}_{\mu(s-2)}':=g^{\mu\mu} {h}_{\mu(s)}$ of  the field
${h}_{\mu(s)}$:
\be
{h}_{\mu(s-2)}'
=
 -\frac{1}{s-1}\,\phi_{2\mu(s-2)} \,.
\ee
The inverse transformations to \p{Phi} have the form
\be
\label{f}
\phi_{\mu(s)}={h}_{\mu(s)}
-\frac{s-1}{4}\,g_{\mu(2)}\,{h}_{\mu(s-2)}'\,,
\qquad
\phi_{2\mu(s-2)}=
-(s-1){h}_{\mu(s-2)}' \,.
\ee
Substituting \eqref{f} into \eqref{LagrF2}
we finally get the Fronsdal Lagangian in $4D$ case:
\begin{eqnarray}
\mathcal{L}_s
&=&
{h}^{\mu(s)}\bigl[\nabla^2+\kappa(s^2-2s-2)\bigr]{h}_{\mu(s)}
-s\,{h}^{\mu(s)}\nabla_\mu\nabla^\lambda{h}_{\lambda\mu(s-1)}\nonumber
\\[.5em]
&&{}
+\frac{1}{2}\,s(s-1)\,{h}^{\mu(s)}\nabla_\mu\nabla_\mu{h}_{\mu(s-2)}'
+\frac{1}{2}\,s(s-1)\,{h}_{\mu(s-2)}'\nabla_\lambda\nabla_\rho{h}^{\lambda\rho\mu(s-2)}
\nonumber
\\[.5em]
&&{}
-\frac{1}{2}\,s(s-1)\,{h}^{\prime\mu(s-2)}\bigl[\nabla^2+\kappa(s^2-3)\bigr]{h}_{\mu(s-2)}'
\nonumber
\\[.5em]
&&{}
-\frac{1}{4}\,s(s-1)(s-2)\,{h}^{\prime\mu(s-2)}\nabla_\mu\nabla^\lambda{h}_{\lambda\mu(s-3)}' \,.
\lb{Fr-L}
\end{eqnarray}
Also, using \eqref{GTAdS-t1} in \eqref{Phi} we obtain gauge transformations of the field ${h}_{\mu(s)}$ \eqref{Phi}:
\be
\delta{h}_{\mu(s)}=s\nabla_\mu\lambda_{\mu(s-1)} \,.
\ee
As a consequence of it, the gauge transformations of their trace part are $\delta{h}_{\mu(s-2)}'=2\nabla^\lambda\lambda_{\lambda\mu(s-2)}$.

Thus, we have demonstrated that the Lagrangian \p{LagrAdS}, obtained within of the framework of the
BRST approach, after eliminating the auxiliary field $\varphi_{1\mu(s-1)}$ is rewritten in the form of
Fronsdal Lagrangian \p{Fr-L}. However, note that the BRST construction from the very beginning results in a Lagrangian
\p{LagrAdS-t} containing more auxiliary fields than the Fronsdal Lagrangian and having more general form then Lagrangian \p{Fr-L}.

%%%%%%%%%%%%%%%%%%%%%%%%%%%%%%%%%%%%%%%%%%%%%%
\section{Summary}\label{ack}
%%%%%%%%%%%%%%%%%%%%%%%%%%%%%%%%%%%%%%%%%%%%%%
In this paper, we have given a concise pedagogical review of BRST
approach to the Lagrangian description for the free massless
arbitrary integer spin fields in the four-dimensional  flat and AdS
spaces in terms of spin-tensor fields. The description is based on
the BRST construction, where the Lagrangian is formulated on the
base of the BRST operator acting in the auxiliary Fock space
generated by the creation and annihilation operators with
two-component spinor indices. No off-shell constraints on the fields
and the gauge parameters are used from the very beginning. We have
demonstrated the advantage of working in four dimensions with
spin-tensor fields, where all traces automatically disappear,
compared to conventional tensor fields. We suppose that the Lagrangian
formulation in terms of fields with two-component spinor indices
will be also very useful for Lagrangian formulations of massive
integer higher-spin models and for free fermionic and supersymmetric
higher-spin models and also for constructing the higher-spin interacting
vertices\footnote{Recent using the spin-tensor fields for
Lagrangian formulation of the $4D$ infinite spin field in AdS space
and massive higher spin fields in constant electromagnetic fields
has been given in \cite{BFIK-1} and \cite {DS} respectively. }.

\end{document}